\documentclass[prb,amsmath,amssymb,showpacs,showkeys,twocolumn]{revtex4-1}
\usepackage{graphicx}
\usepackage{dcolumn}
\usepackage{rotating}
\usepackage{amsthm,amsmath,amssymb,amsfonts,bbm}% Math symbols
\usepackage{mathptmx}
\usepackage{mathrsfs}
\usepackage{xcolor}
\usepackage{siunitx}
\usepackage{bm}
\begin{document}

\title{Delocalized low-frequency magnetoplasmon in a two-dimensional electron fluid with cylindrical symmetry}

\author{I. Kostylev}
\affiliation{Quantum Dynamics Unit, Okinawa Institute of Science and Technology (OIST) Graduate University, Onna, 904-0495 Okinawa, Japan}
\author{M. Hatifi} \altaffiliation[Current address: ]{Quantum Machines Unit, Okinawa Institute of Science and Technology (OIST) Graduate University, Onna, 904-0495 Okinawa, Japan}
\affiliation{Quantum Dynamics Unit, Okinawa Institute of Science and Technology (OIST) Graduate University, Onna, 904-0495 Okinawa, Japan}
\author{A. D. Chepelianskii}
\affiliation{LPS, Univ. Paris-Saclay, CNRS, UMR 8502, F-91405, Orsay, France}
\author{D.\textsc{\textsc{}} Konstantinov}
\affiliation{Quantum Dynamics Unit, Okinawa Institute of Science and Technology (OIST) Graduate University, Onna, 904-0495 Okinawa, Japan}

\begin{abstract}
The properties of a two-dimensional (2D) electron system can be drastically altered by a magnetic field applied perpendicular to the 2D plane. In particular, the frequency of its bulk collective excitations becomes gaped at the cyclotron frequency, while the low-frequency localized excitations, the edge magnetoplasmon (EMP), appear near the system's edge. A new type of the delocalized low-frequency excitations, the gradient magnetoplasmon (GMP), was recently shown to exist in a 2D electron system with a linear density gradient that breaks the system's cylindrical symmetry [Phys. Rev. B \textbf{103}, 075420 (2021)]. Like EMP, these new excitations are gapless and chiral, and originate from the classical Hall effect. Here we show that a similar magnetoplasmon mode can exist in a system with strongly-inhomogeneous radial distribution of the electron density that preserves the cylindrical symmetry. This is experimentally demonstrated in a pristine system of surface electrons on liquid helium and is confirmed by a numerical simulation. This result extends the variety of known collective excitations in a 2D charge system and presents electrons on helium as a promising model system for their study.
\end{abstract}

\date{\today}


\maketitle

\section{Introduction}
\label{sec:intro}

Collective excitations in a system of many interacting particles present a problem of general interest in many scientific fields ranging from condensed-matter physics to material science and engineering. While in general the many-body physics of interacting particles is very complicated, there exists model systems where the many-body dynamics can be well understood and confirmed by the experiments. The two-dimensional electron gas (2DEG) in semiconductors and graphene is of a particular importance due to its simple energy structure and emerging quantum phenomena~\cite{AndoRMP1982,TalSurfSci1990,GrodPRL1991,AliePRL1994,JinNatComm2016,JinNatComm2019,RodNat2020}. A system of electrons trapped on the surface of liquid helium presents another model system, which is a unique classical counterpart of the quantum-degenerate 2DEG~\cite{PetPRL1983}. The distribution of electrons over the motional states is described by the Boltzmann statistics, while the Coulomb interaction between electrons impose on the system the properties of an electron fluid or a Wigner crystal~\cite{Monarkha-book}. Novel collective excitations, such as coupled plasmon-ripplon modes~\cite{FishPRL1979} and chiral edge magnetoplasmons \cite{MastPRL1985,GlatPRL1985,LeaPRL1994}, were first observed in this system. The electron-on-helium system is free of static disorder, which allows to account for the scattering processes and dissipation in the system~\cite{ShikJLTP1974}. 

The plasmon excitation in an electron fluid represents the coherent oscillations of the particle density, with the dispersion law defined by the dimensionality of the system. For an infinite two-dimensional (2D) electron system subject to the magnetic field $B$ applied perpendicular to the 2D plane, the standard hydrodynamic approach predicts existence of propagating charge oscillations with a dispersion law gapped at the cyclotron frequency $\omega_c=\sqrt{eB/m_e}$ (here $e>0$ and $m_e$ is the elementary charge and free electron mass, respectively), which sets the lowest limit on the frequency of propagating magnetoplasmons for a given value of $B$. For a confined 2D system another type of magentoplasmon oscillations can be excited near the system's edge, which induces an abrupt change in the electron density. These chiral edge magnetoplasmons (EMPs) are confined close to the edge and propagate along it in the direction determined by the direction of the applied perpendicular magnetic field. The EMP dispersion law is strikingly different from that of the bulk magnetoplasmons. In particular, the frequency is proportional to the inverse magnetic field and the spectrum is gapless. EMP oscillations originate from the classical Hall effect and their dispersion law can be obtained using the continuity equation and assuming that the charge fluctuations are confined within a quasi one dimensional (quasi-1D) strip along the edge ~\cite{ShikJETP1988}

\begin{equation}
\omega_q = \left( \frac{e n_s q}{2\pi\epsilon_0 B} \right) \ln \frac{1}{q a},  
\label{eq:EMP}
\end{equation}  

\noindent where $a$ is the effective width of the strip. For an unscreened system with a sharp edge, the electric field and the current resulting from the charge density oscillations decay inversely proportional to the distance away from the edge, similar to the field of a charged line. Therefore, the effective width $a$ is of the order of $q^{-1}$. For a realistic case of a finite screened electron system, it depends in a complicated manner on the screening length and the electron density profile~\cite{VolkREV1991}. Often, it is chosen as a phenomenological parameter, which is determined from experiments~\cite{PetePRL1991}. For the typical values of $B$ used in the experiments, the EMP frequency of electrons on helium is several orders of magnitude smaller than that of the bulk magntoplasmons. 

It is commonly accepted that these localized quasi-1D magentoplasmons are the only possible propagating low-frequency modes in a 2D system. However, recently it was demonstrated that a linear gradient in the electron density breaks the cylindrical symmetry of the system and results in a new kind of excitation, the gradient magnetoplasmon (GMP), which is in the same frequency range as EMP~\cite{ChepPRB2021}. Like EMP, these chiral magnetoplasmons originate from the classical Hall effect, while the symmetry-breaking density gradient results in the coupling between different angular harmonics of the propagating excitation. Remarkably, GMP is delocalized from the edge and propagates in the bulk of the system with the dispersion law resembling that given by Eq.~\eqref{eq:EMP}. The existence of GMP was experimentally demonstrated in a circular pool of 2D electrons trapped on the surface of liquid helium, which was also confirmed by a numerical simulation~\cite{ChepPRB2021}. 

A general argument towards the existence of GMP in an infinite 2D system can be made as follows. The electron density fluctuations propagating in $y$-direction, $\delta n\propto e^{i(qy-\omega_qt)}$, produce the Hall current $j_x\propto \sigma_{xy}=en_e/B$, where $n_e$ is the average density of electrons. In the presence of a linear density gradient $\lambda=dn_e/dx\neq 0$, the charge accumulates between the minima and maxima of the density fluctuations, thus sustaining the propagation of $\delta n$ in the $y$-direction. In a way, this is similar to the propagation mechanism of EMP along the edge of a finite system, where the charge density undergoes an abrupt change~\cite{ShikJETP1988}. In the finite electron system with circular geometry considered in Ref.~\cite{ChepPRB2021}, the linear density gradient induces propagating angular excitations and, additionally, facilitates coupling between different angular harmonics. The existence of such excitations, which depended on $\lambda$ by tilting the liquid surface with respect to the confining electrodes, was confirmed in the experiment. 

However, the general argument given above also applies for a system with an inhomogenious radial gradient of the density that preserves its cylindrical symmetry. It might be expected that such a gradient would produce propagating angular excitations, except without coupling between different harmonics. To test this prediction, we have constructed an electron system on the surface of liquid helium with a strongly-nonuniform radial gradient of its density induced by the curvature of the top electrode creating a non-uniform electrostatic environment. In addition to the usual quasi-1D EMP modes, the low-frequency magnetoplasmon exhibit a lower-frequency delocalized 2D magnetoplasmon mode with the dispersion law inversely proportional to the applied magnetic field. The existence of such excitation is confirmed by our numerical simulation, which exhibits a very good agreement with the experimental result. We notice that in gated two dimensional electron gas, sharp changes in electrostatic environment have been shown to create additional modes near the gate boundary recently~\cite{volkov2019}. Here we show how a smooth change in the electrostatic environment can induce a delocalized GMP mode.

% We note that gate plasmons 

\section{Results}
\label{sec:result}
  
\subsection{Experimental setup}
\label{subsec:exp}

\begin{figure}[t]
\includegraphics[width=\columnwidth,keepaspectratio]{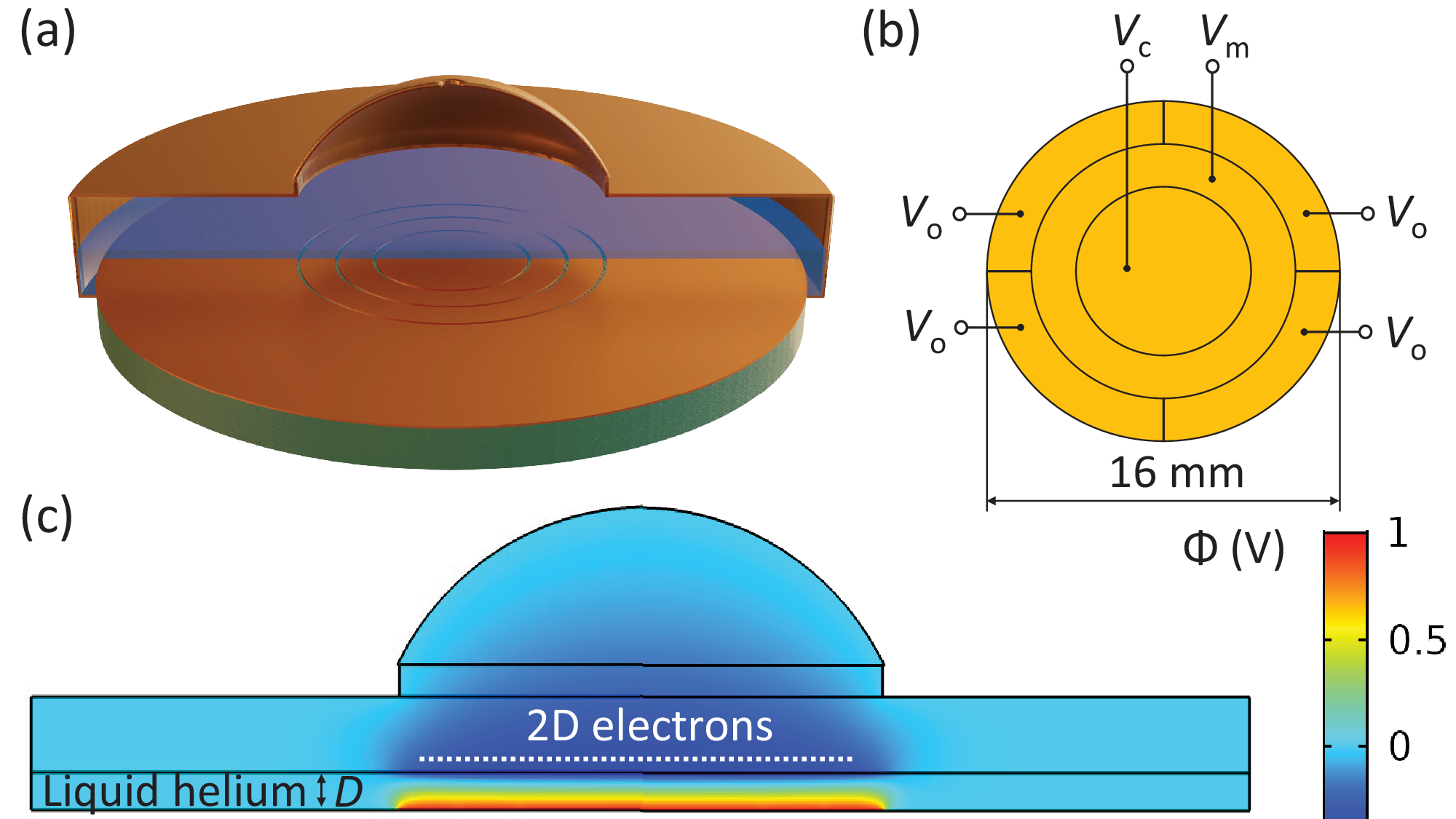}
\caption{\label{fig:1} (color on line) (a) Schematic view of the experimental cell filled with liquid helium. The Corbino disk, used in the experiment to detect and excite charge oscillations in a 2D electron system, is shown at the bottom of the cell, covered by liquid helium. (b) Electrode structure of the Corbino disk. The outer circular electrode is divided into four independent equal-area segments, which break the azimuthal symmetry of the setup. (c) The cross-section of the experimental cell shown in panel (a). The white dot line schematically shows the 2D electron system formed on the surface of liquid  $^3$He set distance $D$ above the cell's bottom. Electrons are accumulated on the free surface of liquid helium above the positively biased Corbino disk, at the bottom of the cell. The color map shows the numerically calculated electric potential $\Phi$ due to a positive bias voltage $V=1$~V applied to the electrodes of the Corbino disk. Note that the electric field above the liquid surface is completely screened for the saturated surface electrons charging density which is shown here.}
\end{figure}

\begin{figure}[t]
\includegraphics[width=\columnwidth,keepaspectratio]{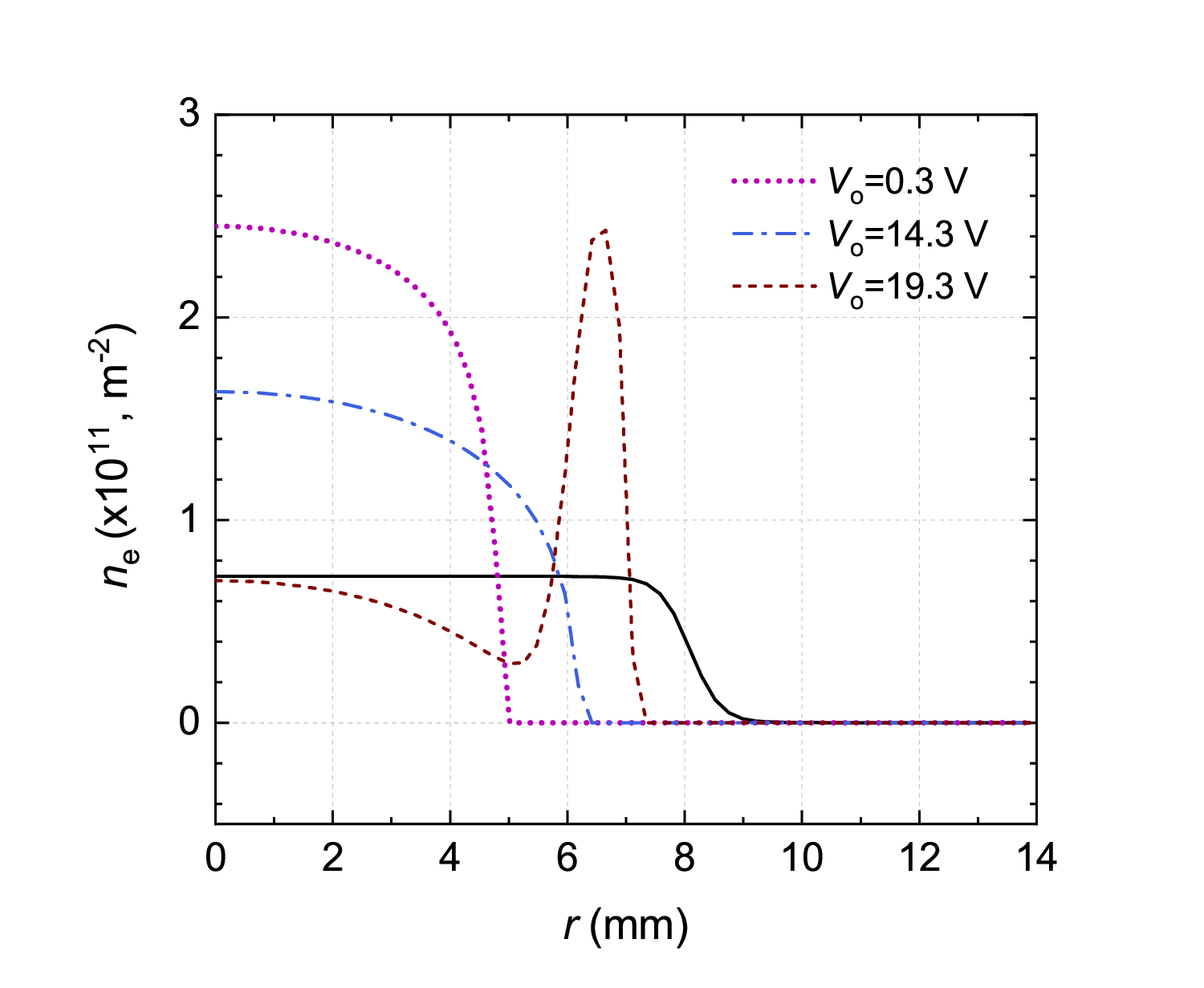}
\caption{\label{fig:2}(color on line)  Radial distribution of the electron density calculated assuming a fixed number of electrons $N_e\approx 14.9\times 10^{6}$ for different voltages  $V_{\rm c}$, $V_{\rm m}$ and $V_{\rm o}$ applied to the electrodes of the Corbino disk. The solid black line corresponds to $V_{\rm c}=V_{\rm m}=V_{\rm o}=1$~V, while the other lines correspond to $V_{\rm c}=V_{\rm m}=14.3$~V and different values of $V_{\rm o}$.}
\end{figure}

In our experiment, the 2D electron system is formed on the free surface of liquid $^3$He condensed in a leak-tight copper experimental cell attached to the mixing chamber of a dilution refrigerator~\cite{KostPRL2021}. The cross section of the cell is shown in Fig.~\ref{fig:1}(a). The cell has the cylindrical symmetry, with a cylindrical lower part and hemispherical upper part. The liquid level is set a distance $D\approx 0.8$~mm above the bottom of the cell. Electrons are generated through thermionic emission from a tungsten filament positioned inside the cell, just above the surface of liquid helium. These electrons accumulate on the liquid's surface above a positively biased round electrode, referred to as the Corbino disk, situated at the center of the cell's bottom. The Corbino disk consists of three concentric electrodes with outer radia 4, 6 and 8~mm separated by two gaps of 0.2~mm width, see Fig.~\ref{fig:1}(b). In addition, the outer electrode is divided into four segments of equal area. Independent dc bias potentials $V_{\rm c}$, $V_{\rm m}$ and $V_{\rm o}$ can be applied to the inner, middle and outer electrodes of the Corbino disk, respectively, as shown schematically in Fig.~\ref{fig:1}(b). Note that in the experiment described here the same potential $V_{\rm o}$ is applied to all four segments of the outer electrode. Initially, the same bias voltage is applied to all three electrodes of the Corbino disk ($V_{\rm c}=V_{\rm m}=V_{\rm o}$) and the surface of liquid helium is charged with electrons. The number of the surface electrons $N_e$ after charging can be determined by calculating the radial density profile of the electron system under the assumption of the complete screening of the electric field above the charged surfaces~\cite{WileJLTP1988}. After the charging, the radial density profile $n_e(r)$ can be varied by independently adjusting the bias potentials $V_{\rm c}$, $V_{\rm m}$ and $V_{\rm o}$. The results of the numerical calculations for $N_e\approx 1.4\times 10^7$, which corresponds to charging the surface with $V=1$~V of the dc potential applied to the Corbino disk electrodes, are shown in Fig.~\ref{fig:2}. The density profile immediately after charging is given by the black solid line corresponding to $V_{\rm c}=V_{\rm m}=V_{\rm o}=1$~V. By increasing the values of $V_{\rm c}=V_{\rm m}$, we can obtain a strongly-nonuniform density profile close to a parabolic shape, e.g. see dotted line in Fig.~\ref{fig:2}. Alternatively, by setting $V_{\rm o}\gtrsim V_{\rm c}=V_{\rm m}$, the density changes nonmonotonically with the radius forming a "caldera" profile, e.g. see short-dashed line. Contrary to the density profiles considered in Ref.~\cite{ChepPRB2021}, in the present experimental setup the density profiles do not show any flat regions for any applied potentials. 

A standard method to excite and detect the charge oscillations in the electron system on liquid helium is to apply an ac driving voltage of amplitude $V_{\rm ac}$ and angular frequency $\omega$ to one of the electrodes of the Corbino disk and then measure the current $I_{\rm ac}$ induced by the electron motion at another electrode (the Sommer-Tanner method)~\cite{ChepPRB2021}. By applying a driving voltage and measuring induced current using the concentric circular electrodes we could probe the radial bulk distribution of the time-dependent charge oscillations, as described below. Alternatively, using the segments of the outer electrode we could probe the angular modes of charge oscillations in the vicinity of  the edge of the electron system~\cite{PetePRL1991,SommPRB1996}. Various parameters, such as the driving frequency, the magnitude of the magnetic field $B$, and the bias voltages $V_{\rm c}$, $V_{\rm m}$ and $V_{\rm o}$, could be varied independently, thus providing information about the dependance of the resonant plasmon frequency on the magnetic field, the size of the system, and the density distribution of electrons in the system. 

\subsection{Experimental observations}
\label{subsec:GMP} 

\begin{figure}[t]
\includegraphics[width=\columnwidth,keepaspectratio]{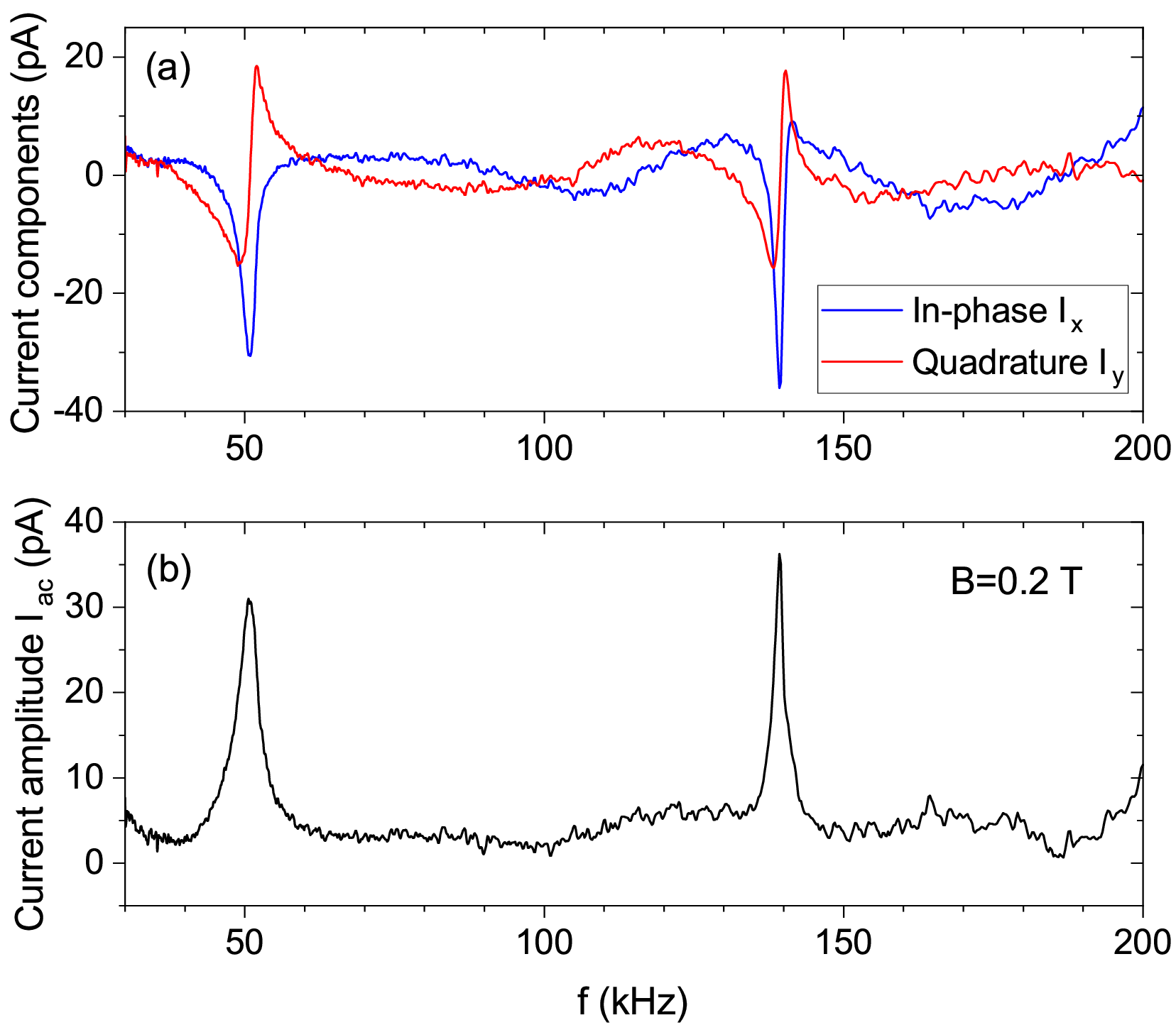}
\caption{\label{fig:3} (color on line) (a) In-phase and quadrature components of the current induced by the motion of electrons at the middle electrode of the Corbino disk when an ac driving voltage at the frequency $\omega$ applied to the center electrode. (b) The amplitude of the current obtained from the current components from panel (a) showing the presence of two resonance modes at 50~kHz and 140~kHz. The simulated density profile for this experiment is shown on Fig.~\ref{fig:2} for $V_o = 14.3\;{\rm V}$.
}
\end{figure}  

\begin{figure}[t]
\includegraphics[width=\columnwidth,keepaspectratio]{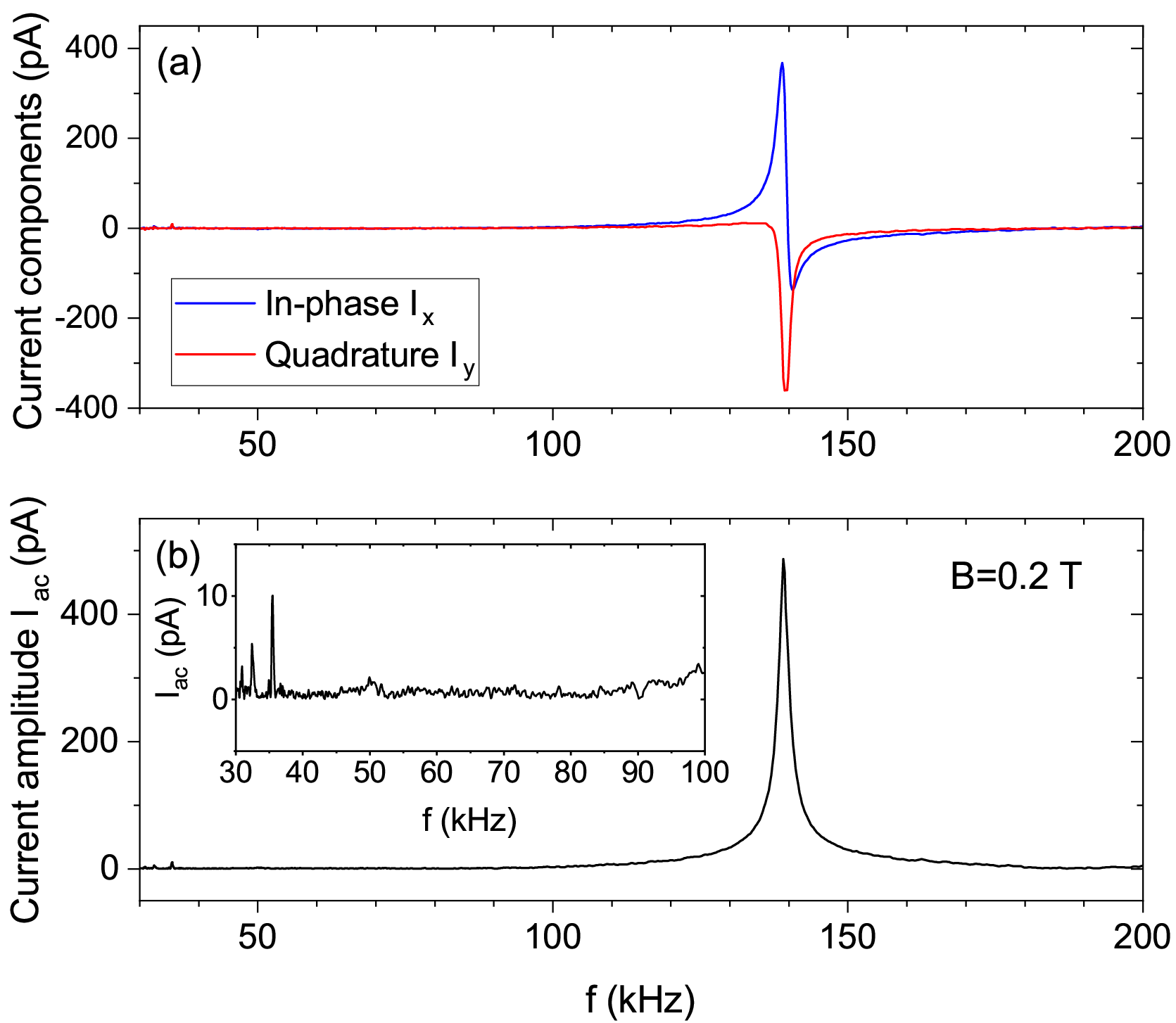}
\caption{\label{fig:4} (color on line) (a) In-phase and quadrature components of the current induced by the motion of electrons at one of the segments of the outer electrode of the Corbino disk when an ac driving voltage at the frequency $\omega$ applied to the center electrode. (b) The amplitude of the current obtained from the data shown in panel (a). The inset shows a magnified signal in the low-frequency range of the main plot. The data are taken at the same magnetic field, temperature and electron density as the data shown in Fig.~\ref{fig:3}.The 50~kHz low frequency resonance is not visible in this figure with detection from segments at the edge of the electron cloud, contrarily to Fig.~\ref{fig:3} where the induced current is measured from the middle electrode. This shows that the 50kHz resonance is not localized on the edge of the system like EMP.  }
\end{figure}

Initially, the surface of liquid helium was charged with an electric bias potential of $1$~V applied to the electrodes of the Corbino disk. After charging, this potential was raised to a higher value in order to confine and stabilize the system against electron loss due to mechanical disturbances. Fig.~\ref{fig:3} shows an example of the magentoplasmon modes detected by the Sommer-Tanner method at the magnetic field $B=0.2$~T, the liquid helium temperature $T=170$~mK and the dc bias potentials $V_{\rm c}=V_{\rm m}=V_{\rm o}=14.3$~V. For such potential, the density of electrons at the center and the radius of the electron system was estimated to be $n_e=1.6\times 10^{11}$~m$^{-2}$ and $R=6$~mm, respectively (see dash-dotted line in Fig.~\ref{fig:2}). The panel (a) shows the in-phase and quadrature components of the electric current induced by electrons at the middle section of the Corbino disk when an ac voltage at the frequency $\omega$ and with the RMS amplitude $V_{\rm ac}=10$~mV is applied to the central section. The cross-talk current due to the driving voltage, which is capacitively coupled to the middle electrode, was subtracted for clarity. The amplitude of the current $I_{\rm ac}$ obtained from the data in Fig.~\ref{fig:3}(a) is shown in Fig.~\ref{fig:3}(b). A similar result was obtained by applying the ac driving voltage with the same frequency and amplitude to the middle electrode, while measuring the induced current at the central electrode. Two clear plasmon resonances are observed at the driving frequency $f\approx 50$ and 140~kHz. For now, we will refer to these two modes as the low-frequency mode and the high-frequency mode, respectively. Note that for $B=0.2$~T used to obtain data shown in Fig.~\ref{fig:4} the cyclotron frequency is $\omega_c/2\pi=5.6$~GHz, thus eliminating the possibility that the observed modes originate from the ordinary bulk magnetoplasmons described by the gapped dispersion law.  

In order to identify these resonant plasmon modes, the current induced by electrons at one of the segments of the outer section of the Corbino disk was measured under the same condition as in Fig.~\ref{fig:3}. Fig.~\ref{fig:4} shows the in-phase and quadrature components (a) and the amplitude (b) of the current signal induced by electrons on a segment of the outer electrode, while the driving ac voltage $V_{\rm ac}=10$~mV is applied to the central electrode. A strong plasmon resonance is observed at the frequency 140~kHz. A series of much weaker noise-like signals of unknown origin is seen at the frequencies below 40 kHz, see the inset of Fig.~\ref{fig:4}(b). Remarkably, the resonant mode at 50 kHz appearing in Fig.~\ref{fig:4} is not observed. The sharp difference between Fig.~\ref{fig:3} and Fig.~\ref{fig:4} clearly shows that the plasmon modes at 50 and 140 kHz have different origin. Again, we confirmed that a similar result to that shown in Fig.~\ref{fig:4} is obtained by applying the ac driving voltage to the middle electrode and measuring the induced current at the same segment of the outer electrode.

\begin{figure}[t]
\includegraphics[width=\columnwidth,keepaspectratio]{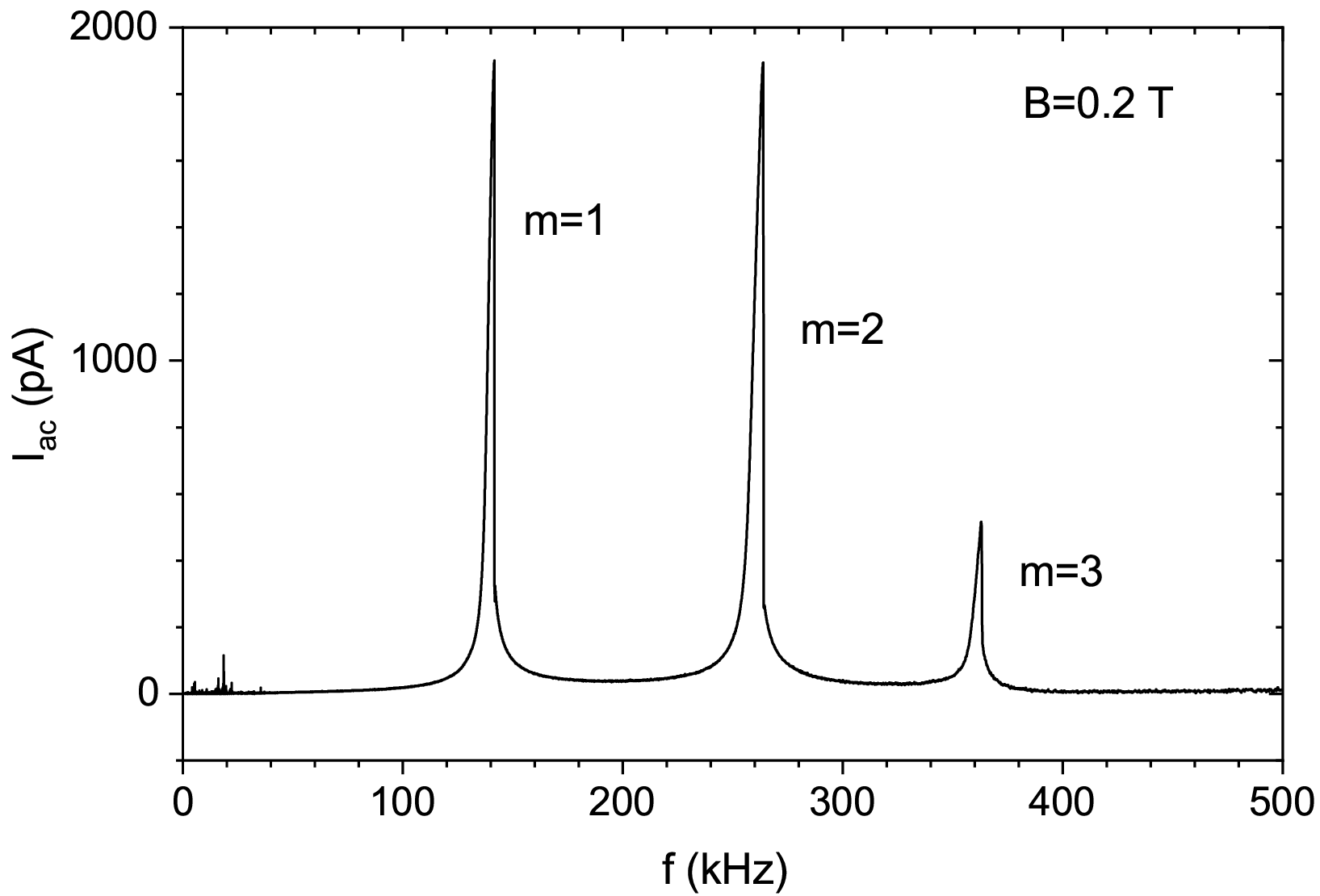}
\caption{\label{fig:5} (color on line) Resonant EMP and its harmonics observed by applying an ac driving voltage to one of the segments of the outer electrode of the Corbino disk, while measuring the amplitude of the induced current $I_{\rm ac}$ at the segment on the opposite side of the outer electrode. The data are taken at the same experimental conditions as the data shown in Figs.~\ref{fig:3} and \ref{fig:4}.}
\end{figure}

The absence of the signal induced by the low-frequency mode at the outer electrode indicates that it corresponds to the bulk distribution of charge oscillations in the electron system. Contrarily, a strongly enhanced signal at 140~kHz detected at the outer electrode indicates that this mode is localized near the edge, that is it corresponds to the ordinary EMP. To verify the latter, we have repeated the experiment at the same magnetic field, temperature and electron density as those used to obtain Figs.~\ref{fig:3} and \ref{fig:4} but applying the ac driving voltage $V_{\rm ac}=10$~mV to one of the segments of the outer electrode of the Corbino disk, while measuring the current $I_{\rm ac}$ induced by the electron motion at another segment located at the opposite side of the outer electrode. We note that this is a standard configuration of the Sommer-Tanner method for detecting the EMP resonances~\cite{SommPhys1994,KiriPRL1995}. The measured current amplitude versus excitation frequency is shown in Fig.~\ref{fig:5}. In addition to the high-frequency mode at $140$~kHz, which can be identified as the fundamental ($m=1$) EMP resonant mode, we observe its higher harmonics ($m=2,3$). As expected, no signatures of the bulk low-frequency mode are observed at 50~kHz. 

\begin{figure}[t]
\includegraphics[width=\columnwidth,keepaspectratio]{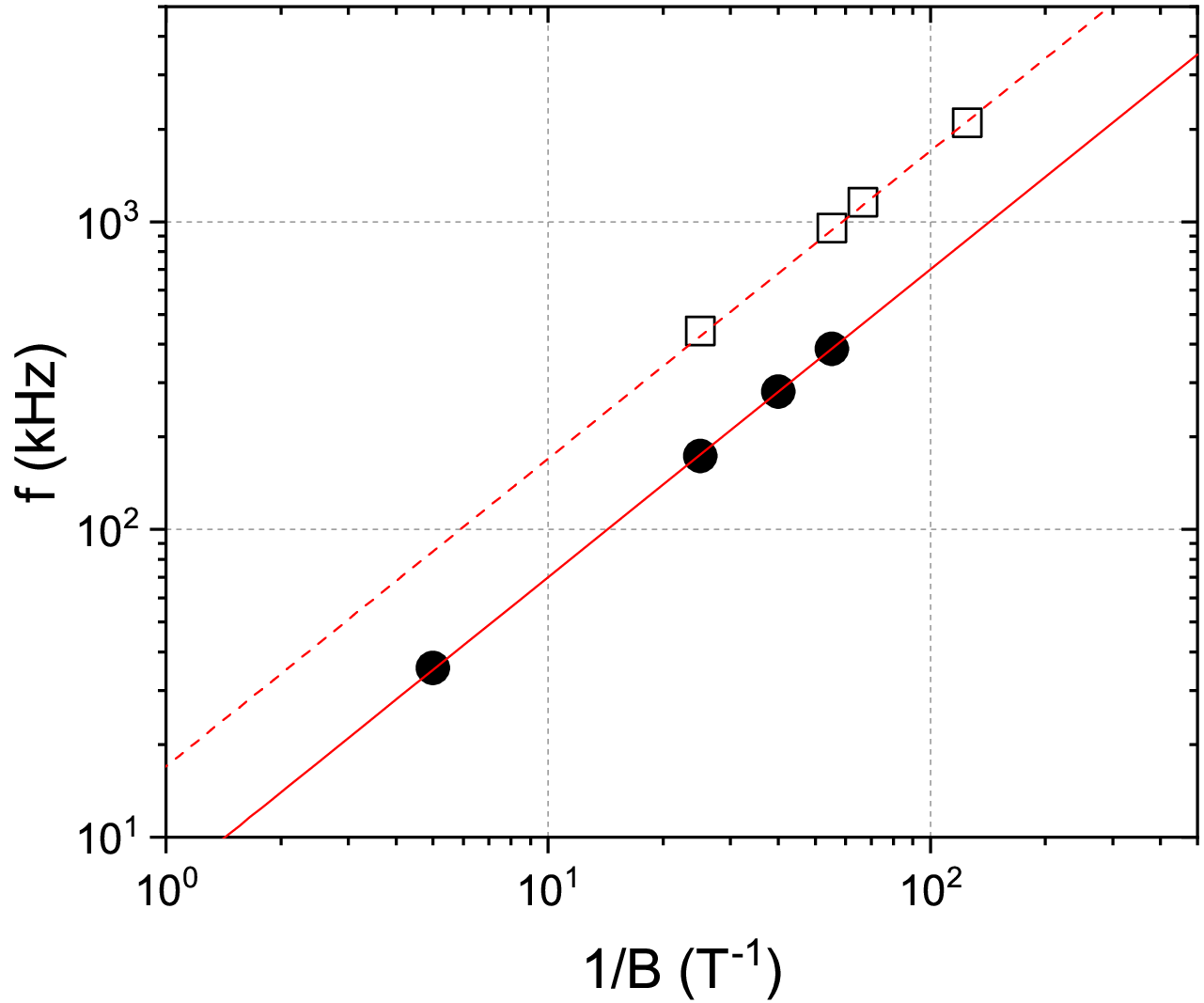}
\caption{\label{fig:6} (color on line) Frequencies of the low-frequency (solid circles) and high-frequency (open squares) magnetoplasmon modes versus the inverse magnetic field. Red lines are linear fits for each mode with the intercept of the y-axis set to zero.}
\end{figure}

It was confirmed that the frequencies of both modes depend linearly on the inverse magnetic field $B$. As an example, Fig.~\ref{fig:6} shown the plot of the resonant frequency versus $B^{-1}$ for the low-frequency (solid circle) and high-frequency (open squares) modes. The data were taken by applying the same dc bias of $14.2$~V to all three electrodes of the Corbino disk. Red lines show linear fits for the low-frequency (solid line) and high-frequency (dashed line) modes with the Y-axis intercept set to zero and with the slope equal to approximately $6.99\pm 0.02$ and $16.97\pm 0.16$~kHz$\cdot$T, respectively. For comparison, the slope estimated from the EMP dispersion law given by Eq.~\eqref{eq:EMP} and neglecting the logarithmic factor is approximately 76.7~kHz$\cdot$T, where we assume $n_e=1.6\times 10^{11}$~m$^{-2}$ and $R=6$~mm, see the dash-dotted line in Fig.~\ref{fig:2}. To estimate the logarithmic factor, we can assume that the effective width $a$ is equal to the screening length $D$ due to the bottom electrode, which results in $\ln (qD)^{-1}\approx 0.18$ and the slope equal to approximately 13.8~kHz$\cdot$T. This agrees reasonably well with the measured slope for EMP in Fig.~\ref{fig:6}.    

\begin{figure}[t]
\includegraphics[width=\columnwidth,keepaspectratio]{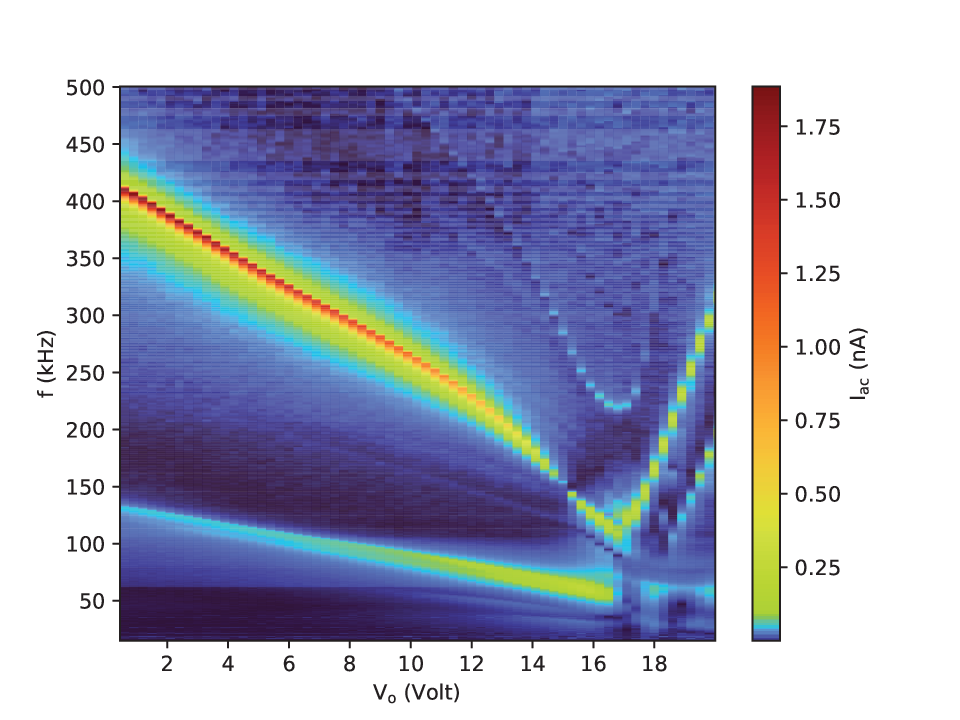}
\caption{\label{fig:7} (color on line) Color map of the current amplitude $I_{\rm ac}$ measured at the middle electrode of the Corbino disk, while an ac driving voltage $V_{\rm ac}=10$~mV is applied to the center electrode, versus the driving frequency $\omega$ and the dc bias voltage $V_{\rm o}$, with fixed $V_{\rm c}=V_{\rm m}=14.3$~V. The data were taken at $B=0.1$~T and $T=170$~mK.}
\end{figure}

Finally, we investigated the dependance of the frequency of the observed modes on the distribution of the electron density. As was described earlier, in our experiment, the radial distribution of the electron density could be varied by applying different dc bias potentials to the three concentric electrodes of the Corbino disk. In particular, by varying the bias potential $V_{\rm o}$ applied to the outer electrode, while fixing the bias potentials applied to the central and middle sections, the radial distribution could be changed as shown in Fig.~\ref{fig:2}. Fig.~\ref{fig:7} shows a color map of the measured current amplitude $I_{\rm ac}$ versus the dc voltage $V_{\rm o}$ and the driving frequency $\omega$. As for the data shown in Fig.~\ref{fig:3}, the driving ac voltage $V_{\rm ac}=10$~mV was applied to the central electrode of the Corbino disk, while the current induced by electrons at the middle electrode was measured. The central and middle electrode bias potential was fixed at $V_{\rm c}=V_{\rm m}=14.3$~V and the data were taken at $B=0.1$~T and $T=170$~mK. At $V_{\rm o}\lesssim 16$~V, the low-frequency mode and the high-frequency EMP mode are clearly observed. At $V_{\rm o}\gtrsim 16$~V, the EMP mode splits into a pair of modes showing the same frequency dependance on $V_{\rm o}$. Note that for such voltages the electron density shows a well-developed "caldera" profile (see Fig.~\ref{fig:2}), therefore it is reasonable to suggest that a pair of resonant modes seen in the experiment at such voltages corresponds to the fundamental EMP modes propagating along the inner and outer edge of electron ring. We note that such interedge-magnetoplasmons (i-EMP) in a ring-shaped electron system on liquid helium has been reported earlier~\cite{SommPRL1995}. This provides an additional support for our identification of the high-frequency mode as a localized EMP.

\subsection{Comparison with numerical simulation}
\label{subsec:simul} 

\begin{figure}[t]
\includegraphics[width=\columnwidth,keepaspectratio]{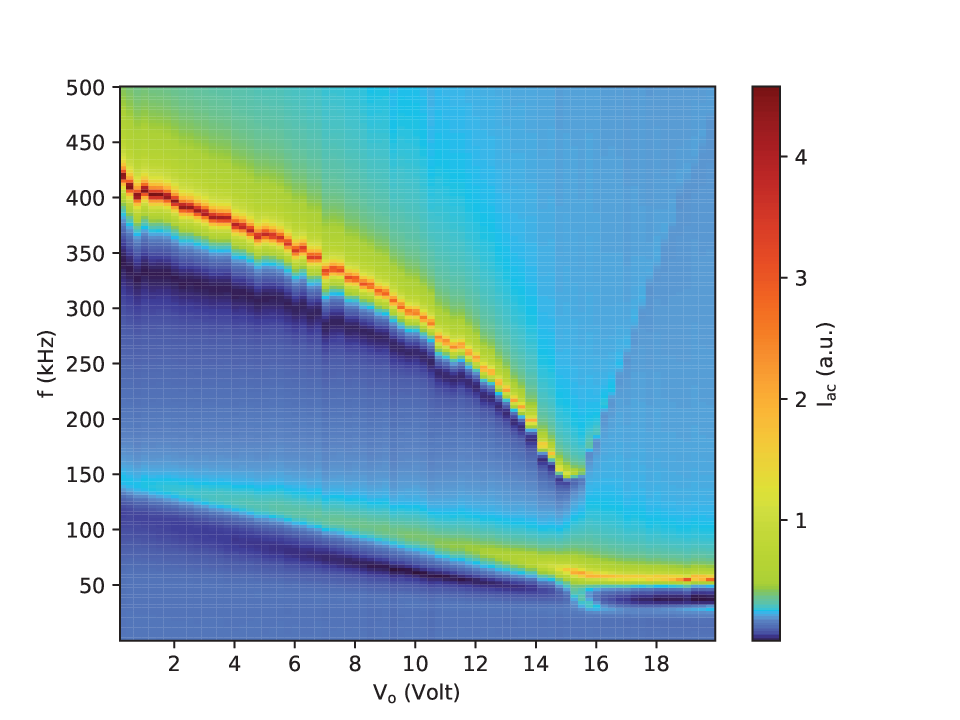}
\caption{\label{fig:8} (color on line) Numerical simulation of the current signal (in arbitrary units) induced at the middle electrode by the electron density oscillations under the same magnetic field and the electron density distribution as in Fig.~\ref{fig:7}.}
\end{figure}

In order to confirm the behavior observed in Fig.~\ref{fig:7}, we carried out a numerical simulation of the current signal induced by the density fluctuations at the middle electrode under the same conditions as in Fig.~\ref{fig:7}(a). The results of the simulation are shown in Fig.~\ref{fig:8}. The simulation assumes $N_e=1.1\times 10^7$ electrons having the radial density distribution preserving the cylindrical symmetry and includes only the fundamental angular harmonic of the density fluctuations. Other details of the numerical method were described previously~\cite{ChepPRB2021}. Overall, we find a rather remarkable agreement between the experimental data and the simulation results. In particular, the delocalized low-frequency mode is clearly reproduced by our simulation.                 

\section{Discussion}
\label{sec:disc}

In the experiment reported here we clearly identified two different modes which have an identical dependance of their frequencies on the magnetic field $B$ (see Fig.~\ref{fig:6}) but a different behavior depending on the driving condition and configuration of the electron system, e.g. see Fig.~\ref{fig:7}. It follows from the earlier discussion that both modes originate from the same physical mechanism, namely the classical Hall effect, which explains the inverse $B$-dependance of the dispersion law. Both modes correspond to the angular standing wave excitation, as is confirmed by our numerical simulation. The standard electrode configuration which allows to excite and detect the angular modes of EMP is a segmented electrode which breaks the  cylindrical symmetry of the measuring setup~\cite{SommPhys1994,KiriPRL1995}. It is clear that the angular modes can not be excited and detected by a pair of circular electrodes that preserves the cylindrical symmetry, which is opposite to what is observed, for example, in Fig.~\ref{fig:3}. However, note that in a real experimental setup there are always some deviations from the cylindrical symmetry due to a misalignment of the top and bottom parts of the experimental cell, a tilt of the liquid helium surface with respect to the cell electrodes, etc. Such deviations can cause the excitation of the angular modes of the charge oscillations by applying the voltage drive to the circular electrodes. We note that in the experiment reported here the frequency of the observed low-frequency mode was found to be independent on the tilting angle in the range $\pm 0.5^{\circ}$. This agrees with our expectation that this mode originates from the strongly-nonuniform radial density profile of the electron system induced by the curved top electrodes rather than from the linear density gradient considered previously~\cite{ChepPRB2021}. However, our numerical simulations suggest that the low frequency GMP mode reported here smoothly evolves into the GMP mode induced by a linear density gradient when its amplitude increases (see Appendix). Confirming this connection between GMP modes requires further study.

\section{Conclusions}
\label{sec:conclude}

To summarize, we showed that a new type of delocalized low-frequency magnetoplasmon can exist in a confined electron system with strongly-nonuniform radial density distribution which preserves the cylindrical symmetry of the system. The existence of such magnetoplasmon is experimentally demonstrated in a clean system of electrons trapped on the surface of liquid helium and is confirmed by the results of a numerical simulation. Realization of such a magnetoplasmon in other mesoscopic systems, such as 2DEG in heterostructures, could be of a particular interest. In general, we believe that this work expands our knowledge of the collective phenomena in charged systems and demonstrates the system of electrons on helium as a promising platform for their study.

{\bf Acknowledgements} This work is supported by the internal grant from the Okinawa Institute of Science and Technology (OIST) Graduate University.

$   $
\appendix 
\section{simulations with a linear density gradient}

We investigated numerically the influence of a linear density gradient on the frequency of the GMP mode in the experimental cell geometry with electrostatic inhomogenity induced by curved top electrodes. For this purpose we simulated the effect of a constant density gradient $\lambda$ using equations from \cite{ChepPRB2021} and taking into account the curved electrode geometry. For $\lambda = 0$, the shown data corresponds to a slice of Fig.~\ref{fig:8} for $V_{\rm o} = 5\;{\rm V}$. Since there is no coupling between angular harmonics in this simulation, we assumed an artificial asymmetric excitation so that the EMP mode can be excited. For $\lambda > 0$ we simulated the response expected from central excitation as in the experiment assuming that $\lambda$ is the only coupling term between harmonics. For small $\lambda = 5\times10^5\;{\rm cm^{-3}}$, the resonance frequency is not changed and we do not see any additional GMP mode due to the gradient, while for higher values of $\lambda = 10^6\;{\rm cm^{-3}}$ both GMP and EMP peaks are shifting due to the gradient. This suggests that there is a single low frequency GMP mode whose frequency is given by the main source of non-uniformity (curved electrodes or density gradients). However more detailed theoretical investigations are needed to settle this question reliably. We notice also that the simulations with finite $\lambda$ predict a weaker response of the EMP compared to GMP for a priori symmetric excitation on the central electrode. This is also what we observed for the flat electrodes geometry previously~\cite{ChepPRB2021}. In the present experiment, we observe a similar EMP and GMP response amplitude. This probably indicates an additional source of symmetry breaking, such as a misalignement between the centers of top and bottom electrodes. Further investigations are needed to clarify this.

\begin{figure}[h]
% figure and figure data in: 
% ~/Manip/diluhelium/VisiteDenisJan2016/FreeFem/simu2023
\includegraphics[width=\columnwidth,keepaspectratio]{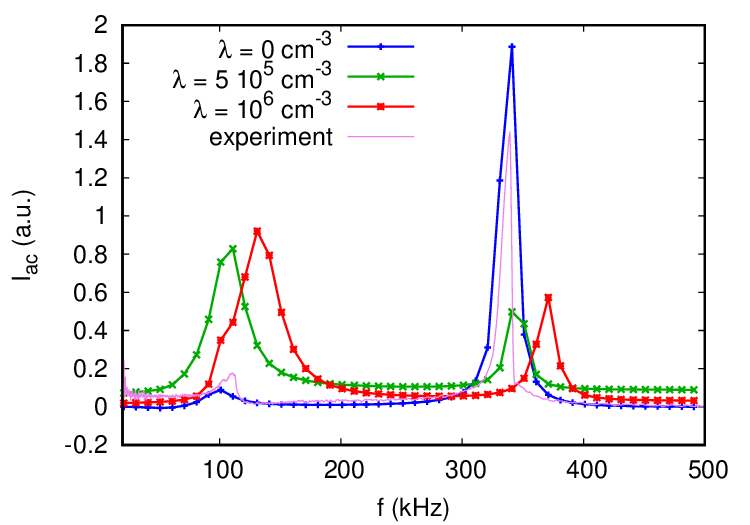}
\caption{\label{fig:a1} (color on line) Simulations of the effect of an uniform density gradient on the frequency of EMP and GMP modes.}
\end{figure}


\section*{References}

\begin{thebibliography}{99}

\bibitem{AndoRMP1982} T. Ando, A. B. Fowler, and F. Stern, Rev. Mod. Phys. \textbf{54}, 437 (1982).

\bibitem{TalSurfSci1990} V. K. Talyanskii, M. Wassermeier, A. Wixforth, J. Oshinowo, J. P. Kotthaus, I. E. Batov, G. Weimann, H. Nickel, and W. Schlapp, Surf. Sci. \textbf{229}, 40 (1990).

\bibitem{GrodPRL1991} I. Grodnesky, D. Heitmann, and K. von Klitzing, Phys. Rev. Lett. \textbf{67}, 1019 (1991).

\bibitem{AliePRL1994} I. L. Aleiner and L. I. Glazman, Phys. Rev. Lett. \textbf{72}, 2935 (1994).

\bibitem{JinNatComm2016} D. Jin, L. Lu, Z. Wang, C. Fang, J. D. Joannopolos, M. Sojjacic, L. Fu, and N. X. Fang, Nat. Comm. \textbf{7}, 13486 (2016).

\bibitem{JinNatComm2019} D. Jin, Y. Xia, T. Christensen, M. Freeman, S. Wang, K. Y. Fong, G. C. Gardner, S. Fallahi, Q. Hu, Y. Wang, L. Engel, Z.-L. Xiao, M. J. Manfra, N. X. Fang, and X. Zhang, Nat. Commun. \textbf{10}, 4565 (2019). 

\bibitem{RodNat2020} A. Rodin, M. Trushin, A. Carvalho, and A. H. Castro Neto, Nat. Phys. Rev. \textbf{2}, 524 (2020).  

\bibitem{PetPRL1983} F. M. Peters and P. M. Platzman, Phys. Rev. Lett. \textbf{50}, 2021 (1983). 

\bibitem{Monarkha-book} Y. P. Monarkha and K. Kono {\it Two-Dimensional Coulomb Liquid and Solids} (Springer, Berlin, 2004).

%\bibitem{PinePR1952} D. Pine and D. Bohm, Phys. rev. \textbf{85}, 338 (1952).

\bibitem{FishPRL1979} D. Fisher, B. I. Halperin, and Pm. M. Platzman, Phys. Rev. Lett. \textbf{42}, 798 (1979).

\bibitem{MastPRL1985} D. B. Mast, A. J. Farm, and A. L. Fetter, Phys. Rev. Lett. \textbf{54}, 1706 (1985).

\bibitem{GlatPRL1985} D. C. Glattli, E. Y. Andrei, G. Deville, J. Poitrenaud, and F. I. B. Williams, Phys. Rev. Lett. \textbf{54}, 1710 (1985).

\bibitem{LeaPRL1994} M. J. Lean, P. Fazooni, P. J. Richardson, and A. Blackburn, Phys. Rev. Lett. \textbf{73}, 1142 (1994).

\bibitem{ShikJLTP1974} V. B. Shikin and Yu. P. Monarkha, J. Low Temp. Phys. \textbf{112}, 193 (1974).

\bibitem{ShikJETP1988} V. B. Shikin, JETP Lett. \textbf{47}, 555 (1988).

\bibitem{VolkREV1991} V. A. Volkov and S. A. Mikhailov, {\it Landau Level Spectroscopy, Modern Problems in Condensed Matter Science}, Vol. 27 (1991), Chap. 15.

\bibitem{PetePRL1991} P. J. Peters, M. J. Lea, A. M. Janssen, A. O. Stone, W. P. N. Jacobs, P. Fozooni, and R. W. van der Heijden, Phys. Rev. Lett. \textbf{67}, 2199 (1991).

\bibitem{ChepPRB2021} A. D. Chepelinaskii, D. Papoular, D. Konstantinov, H. Bouchiat, and K. Kono, Phys. Rev. B \textbf{103}, 075420 (2021).

\bibitem{KostPRL2021} I. Kostylev, A. A. Zadoroshko, M. Hatifi, and D. Konstantinov, Phys. Rev. Lett. \textbf{127}, 186801 (2021).

\bibitem{volkov2019} A. A. Zabolotnykh and V. A. Volkov, Phys. Rev. B {\bf 99}, 165304 (2019)
  
\bibitem{WileJLTP1988} L. Wilen and R. Giannetta, J. Low Temp. Phys. \textbf{72}, 353 (1988). 

\bibitem{SommPRB1996} P. K. H. Sommerfeld, R. W. van der Haijden, and F. M. Peeters, Phys. Rev. B \textbf{53}, R12250 (1996).

\bibitem{ShirJLTP1995} K. Shirahama, S. Ito, H. Suto and K. Kono, J. Low Temp. Phys. \textbf{101}, 439 (1995).

\bibitem{SommPhys1994} P. K. H. Sommerfeld, R F. M. Peeters, H. F. W. J. Vorstenbosch, R. W. van der Haijden, A. T. A. M. de Waele, and M. J. Lea, Physica B \textbf{194}, 1311 (1994).

\bibitem{KiriPRL1995} O. I. Kirichek, P. K. H. Sommerfeld, Yu. P. Monarkha, R F. M. Peeters, Yu. Z. Kovdrya, P. P. Steijaert, R. W. van der Haijden, and A. T. A. M. de Waele, Phys. Rev. Lett. \textbf{195}, 1190 (1995).

\bibitem{SommPRL1995} P. K. H. Sommerfeld, P. P. Steijaert, R F. M. Peeters, and R. W. van der Haijden, Phys. Rev. Lett. \textbf{74}, 2559 (1995).

\end{thebibliography}
\end{document}